\documentclass[conference,a4paper]{IEEEtran}
\usepackage{xcolor}
\usepackage{balance}
\usepackage{booktabs}
\usepackage{amsfonts,amssymb,amsmath}
\usepackage{url}
\usepackage{algorithmic}
\usepackage{placeins} 
\usepackage{array} 
\usepackage{multirow}

\ifCLASSINFOpdf
  \usepackage[pdftex]{graphicx}
\else
  \usepackage[dvips]{graphicx}
\fi
\ifCLASSOPTIONcompsoc
 \usepackage[caption=false,font=normalsize,labelfont=sf,textfont=sf]{subfig}
\else
 \usepackage[caption=false,font=footnotesize]{subfig}
\fi
\begin{document}
%
\title{Centimeter-level Geometry Reconstruction and Material Identification in 300~GHz\\ Monostatic Sensing}

\author{Zitong Fang\IEEEauthorrefmark{1}, 
Ziming Yu\IEEEauthorrefmark{2}, 
and Chong Han\IEEEauthorrefmark{1} \\

\IEEEauthorblockA{\IEEEauthorrefmark{1}Terahertz Wireless Communications (TWC) Laboratory, Shanghai Jiao Tong University, China \\ Email:  \{zitong.fang,chong.han\}@sjtu.edu.cn\\ 
\IEEEauthorrefmark{2}Huawei Technologies Co., Ltd, China. Email: yuziming@huawei.com \\
}
}



\maketitle

\begin{abstract}
Terahertz (THz) integrated sensing and communication (ISAC) technology is envisioned to achieve high communication performance alongside advanced sensing abilities. For various applications of ISAC, accurate environment reconstruction including geometry reconstruction and material identification is critical. This paper presents a highly precise geometry reconstruction algorithm and material identification scheme for a monostatic sensing case in a typical indoor scenario. Experiments are conducted in the frequency range from 290~GHz to 310~GHz using a vector network analyzer (VNA)-based channel sounder by co-locating the transmitter and receiver. A joint delay and angle space-alternating generalized expectation-maximization (SAGE)-based algorithm is implemented to estimate multipath component (MPC) parameters and the indoor geometry is reconstructed based on the extracted parameters. Furthermore, a geometry-based method is employed to model and remove the spurious path of the corner, reaching an accuracy of 1.75~cm. Additionally, a material database using THz time-domain spectroscopy (THz-TDS) is established, capturing reflection losses of over 200 common material samples. Applying this database to our monostatic sensing, the measured reflection losses of wall and window frame are accurately identified as cement and steel, respectively. Our results demonstrate the centimeter-level geometry reconstruction and accurate material identification for practical THz ISAC scenarios, which unleash unprecedented sensing potential compared to microwave and millimeter-wave bands.
\end{abstract}


%

\section{Introduction}



The next-generation mobile networks are expected to offer significant communication and sensing abilities. Therefore the integration of sensing and communication (ISAC) is foreseen to be one of the key technologies. Through the application of the Terahertz (THz) band, which offers unexplored spectrum resource and enables high distance resolution~\cite{han2024thz}, the THz ISAC system can support devices to gain comprehensive awareness of their surroundings and enhance the sensing capabilities~\cite{dupleich2024characterization,wu2021thz}. For future ISAC applications such as smart industry and smart transportation, accurate environment reconstruction is critical. In particular, the technique of geometry reconstruction and material identification together provide a comprehensive understanding of the surrounding environment.

The incorporation of THz communication and sensing facilitates high-resolution geometry reconstruction by exploiting the wide bandwidth in the THz band. A handful of state-of-the-art measurement-based THz ISAC research has been conducted. For instance, the authors in~\cite{li2023300} conducted THz mapping experiments in three indoor scenarios, where an average error of 0.1~m between the reconstructed geometry and the ground truth is observed. The authors in~\cite{lotti2023radio} proposed a radio simultaneous localization and mapping (SLAM) algorithm for indoor scenarios to derive the map of laboratory/office and infer users' trajectories. However, despite these advancements, current geometry reconstruction methods achieve decimeter-level accuracy that are limited to degraded precision in complex regions such as corners. 

Furthermore, material identification for environment reconstruction is essential in ISAC systems. Different materials exhibit unique electromagnetic (EM) properties, particularly in the THz band. Previous studies have utilized THz time-domain spectroscopy (THz-TDS) to investigate materials, measuring reflection properties from 100 to 500~GHz~\cite{jansen2008impact}. 
Additionally, angle- and frequency-dependent
measurements, coupled with Kirchhoff scattering theory, have
been used to model rough surface materials from 0.1 to 1~THz~\cite{jansen2011diffuse}. The absorption and reflection characteristics of typical building materials at various angles from 100 to 350~GHz are measured~\cite{kleine2005characterization}. Although some efforts have been made to characterize material properties in the ISAC system using THz-TDS, practical measurements for material identification of real environment reconstruction are still lacking. 

Existing works focus on either channel measurements or material database based on THz-TDS, while we tackle both geometry and material sensing that enhance environment reconstruction. In this paper, monostatic sensing experiments are conducted in a laboratory using a vector network analyzer (VNA)-based sounder, using a 20~GHz bandwidth from 290~GHz to 310~GHz. The transmitter (Tx) and receiver (Rx) are co-located on a rotator which can perceive the surrounding environments from all directions. First, we present a high resolution parameter estimation (HPRE) estimator based on a joint delay and angle space-alternating generalized expectation-maximization (SAGE) algorithm that improves geometry reconstruction accuracy. Additionally, a geometry-based method is employed to remove the corner effect of spurious path. Second, we employ THz-TDS to establish a material database including over 200 common material samples. To demonstrate, the reflection losses of two typical materials calculated from the VNA-based measurements are then compared with the database, and the materials of the wall and window frame are identified as cement and steel, respectively. This work attaches great importance to high indoor environment geometry reconstruction accuracy of centimeter-level and joint material identification based on THz-TDS. 

The remainder of the paper is organized as follows. The monostatic sensing experiments are presented in Sec. II. In Sec. III, the signal model and data processing algorithm are explained. Experiment results are calculated and analyzed in Sec. IV. Finally, Sec. V concludes the paper.

\section{VNA and TDS Measurement Systems}
In this section, the monostatic sensing experiments are described, including the sounding system, measurement scenario, and deployment for material identification.
\subsection{Monostatic Sensing Measurement Based on VNA}
\begin{figure}
\centering
\subfloat[THz measurement system and setup.]{\includegraphics[width=0.75\columnwidth]{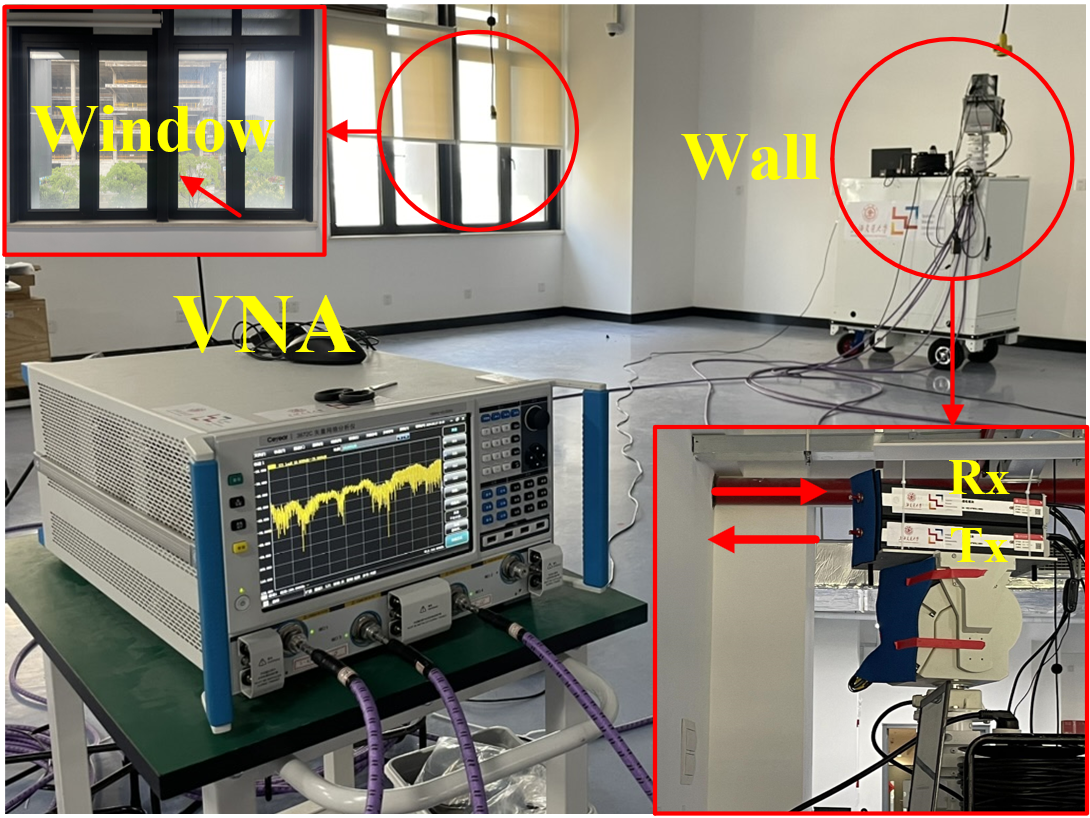}%
\label{fig:scenario1}}
\hfil
\subfloat[Schematic of the scenario.]{\includegraphics[width=0.75\columnwidth]{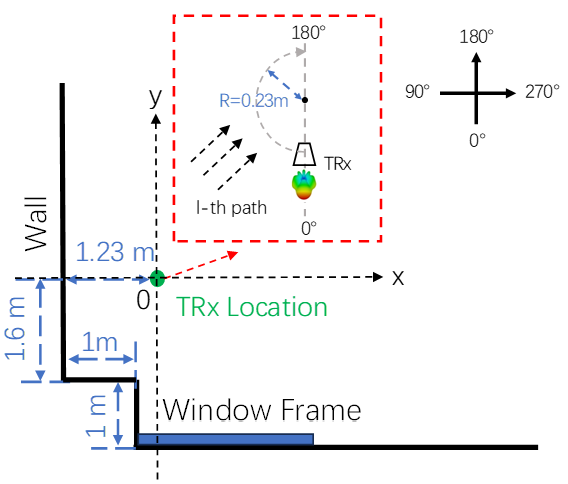}%
\label{fig:scenario}}
\caption{Measurement system and scenario.}
\label{fig:description}
\end{figure}

The monostatic sensing measurements are conducted in an empty area in a laboratory scenario, as shown in Fig.~\ref{fig:description}. In these measurement campaigns, a VNA-based channel sounder is employed in~\cite{10494205,lyu2021design} with a frequency range of $290$-$310$~GHz and $2001$ frequency points. The directional scanning scheme (DSS) is applied to capture the channel spatial profiles. As shown in Fig.~\ref{fig:description}(a), the Tx and Rx are mounted on the same rotator, mimicking a transceiver (TRx), which investigates the THz wave interaction with the common indoor structures. Specifically, a uniform circular array (UCA) configuration is employed, in which the antennas are positioned in a circular layout with a radius R = 0.23 m. The UCA is designed to provide uniform angular coverage, enabling high-resolution sensing across the azimuthal plane.

During the experiment, a 20~GHz wide band is measured, which corresponds to a delay resolution of 0.05~ns and a distance resolution of 1.5~cm. The height of the TRx is 2.0~m. Moreover, both Tx/Rx are equipped with horn antennas, whose gain and half-power beamwidth (HPBW) are 26~dBi and $8^{\circ}$, respectively. As depicted in Fig.~\ref{fig:description}(b), to capture environment information from different directions, the TRx scans in the azimuth plane with $1^{\circ}$ angle step, from $0^{\circ}$ to $180^{\circ}$. The experiment configurations are listed in Table~\ref{tab:Experiment parameters}. 

\begin{table}[]
\caption{Experiment configurations}
\label{tab:Experiment parameters}
\centering
\begin{tabular}{cc}
\toprule
Parameter              & Values              \\
\midrule
Frequency band         & 290-310~GHz          \\
Bandwidth              & 20~GHz               \\
Delay resolution       & 0.05~ns             \\
Distance resolution & 1.5~cm               \\
TRx height             & 2.0~m                \\
Antenna gain of Tx/Rx  & 26~dBi               \\
HPBW of Tx/Rx antennas & $8^{\circ}$                  \\
Azimuth rotation range & {[}$0^{\circ}$:$1^{\circ}$:$180^{\circ}${]} \\
\bottomrule
\end{tabular}

\end{table}
\begin{figure}[t]
\centering
\includegraphics[width=0.55\columnwidth]{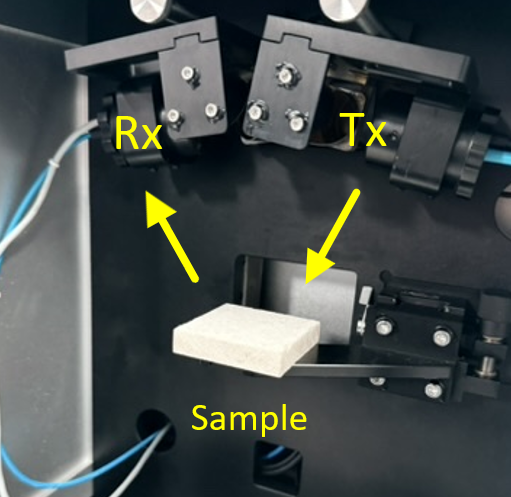}
\caption{THz-TDS system.}
\label{fig:TDS}
\end{figure}

\subsection{THz-TDS System}
In this paper, THz-TDS is employed to identify materials by measuring the reflection loss ranging from 0 to 6~THz~\cite{menlo2024terahertz}, and we use only the results in 300~GHz to compare with our measurements. Fig.~\ref{fig:TDS} shows the employed THz-TDS system. Both the incident and reflection angles of the Tx and Rx antennas are set at $15^{\circ}$ to ensure a consistent measurement of reflection properties. The reflection loss at this angle is expected to be close to $0^{\circ}$. The sample under test is positioned at a fixed 9~cm from the antennas. Note that the free-space path loss (FSPL) is compensated. The experiments are conducted with a wide variety of common materials, which are categorized into metals, biological materials, building materials, and functional materials. The test set includes over 200 material samples and is classified into more than 30 types. The reflection loss of each sample can be calculated after obtaining the received signal strength.

\section{Data Processing for Environment Reconstruction}
In this section, the data processing procedures for environment reconstruction are introduced. To begin with, the measured channel frequency responses are preprocessed. Furthermore, the multipath component (MPC) parameters are estimated using a two-dimensional (2D) SAGE algorithm, whose parameters are utilized for mapping. Additionally, a geometry-based method is employed to remove the corner effect. 
\subsection{Data Preprocessing}
\begin{figure}[t]
\centering
\includegraphics[width=0.85\columnwidth]{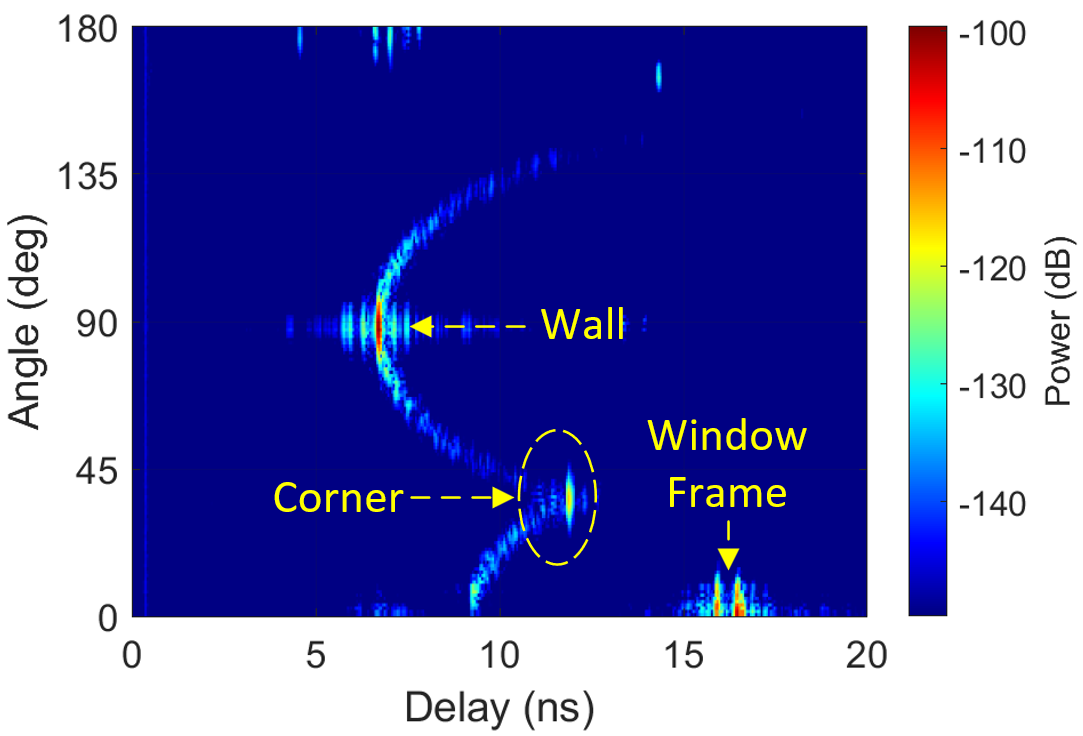}
\caption{PADP for the measurement at TRx Location.}
\label{fig:PADP}
\end{figure}

Through the measurement results of VNA, the channel frequency response is extracted ranging from 290~GHz to 310~GHz. The frequency-domain data is first calibrated and then transformed into the delay domain through the inverse discrete Fourier transform (IDFT) to obtain the channel impulse response (CIR)~\cite{li2023300}. 

Specifically, the power-angle-delay profile (PADP) is calculated by
\begin{equation}
P(\tau, \varphi) = 20 \log_{10} \left( |h(\tau, \varphi)| \right),
\end{equation}
where $h(\tau,\varphi)$ denotes the received CIR, $\tau$ represents the propagation delay and $\varphi$ denotes the rotation angle of UCA. As shown in Fig~\ref{fig:PADP}, the resulting PADP provides a detailed representation of the MPC characteristics, capturing the interactions between the THz signal and the environment. Three key structures can be clearly observed as wall, corner, and window frame. The wall is characterized by a continuous reflection, and the corner presents a strong specular path. Additionally, the window frame is identifiable with a strong reflected signal due to its metallic composition.

\subsection{HPRE Estimator}
To estimate MPCs in the environment, a joint delay and angle SAGE algorithm is adopted. The received signal model can be defined as
\begin{equation}
\label{equ:signal}
H(f, \varphi) = \sum_{\ell=1}^{L} \alpha_{\ell} e^{-j 2 \pi f \tau_{\ell}} \cdot a_{\rm TRx,\ell}(f, \varphi) + W(f,\varphi),
\end{equation}
where $L$ is the total number of MPCs, $f$ is the carrier frequency, and $W(f,\varphi)$ represents the noise. $\alpha_{\ell}, \tau_{\ell}$ denote the path gain and delay of the $\ell$-th MPC, respectively. $\Theta = \{\alpha_{\ell}, \tau_{\ell}, \theta_{\ell}\}$ represents the estimated MPC parameters, where $\theta_{\ell}$ denotes the angle of the $\ell$-th MPC. Note that since the Tx and Rx are co-located on the same rotator, the angle represents both the transmit and receive angles of the $\ell$-th MPC, i.e., $\theta_{\ell}=\theta_{Tx,\ell}=\theta_{Rx,\ell}$. The antenna array response of TRx is given by
\begin{equation}
a_{\rm TRx,\ell}(f, \varphi) = e^{j 4 \pi f R \cos(\theta_{\ell} - \varphi)/c} \cdot G_{\rm TRx}(\theta_{\ell} - \varphi),
\end{equation}
where $c$ is the speed of light, $R$ denotes the radius of UCA. The term $G_{\rm TRx}(\theta_{\ell} - \varphi)$ represents the total antenna gain of TRx at the angle $\theta_{\ell}$.

We adopt a 2D SAGE algorithm that simultaneously estimates the delay and angle of MPCs, and the likelihood function is iteratively maximized to refine the estimates of the parameters for each MPC~\cite{fessler1994space,fleury1999channel}. The estimated MPCs parameters are given by
\begin{equation}
\hat{\Theta} = \arg\max_{\Theta} 2 \text{Re}\left\{ \mathbf{H_{r}}(f, \Theta) \cdot s_{\ell}^*(f, \Theta) \right\} - |s_{\ell}(f, \Theta)|^2,
\end{equation}
where $\mathbf{H_{r}}(f, \Theta)$ is the received signal and $s_{\ell}(f, \Theta)$ is the reconstructed signal using the aforementioned signal model, i.e.,
\begin{equation}
    s_{\ell}(f, \Theta)=\alpha_{\ell} e^{-j 2 \pi f \tau_{\ell}} \cdot a_{\rm TRx,\ell}(f, \varphi).
\end{equation}
\( \text{Re}\{ \cdot \} \) denotes taking the real part of a complex number, \( \cdot^* \) represents the complex conjugate, and \( |\cdot| \) denotes the modulus of a complex number.

By incorporating both delay and angle in the estimation process as well as considering the antenna radiation pattern from the measurement, this 2D SAGE algorithm allows for high resolution in environment reconstruction. 
\subsection{Data Processing For Mapping}
\label{sec:mapping}
\begin{figure*}[t]
\centering
\subfloat[Maximum-search algorithm.]{\includegraphics[width=0.68\columnwidth]{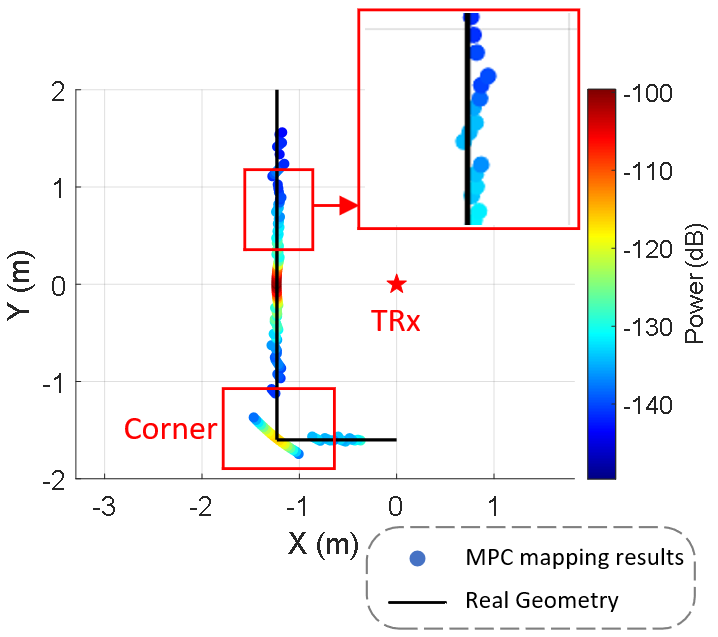}%
\label{fig:recon_maxp}}
\hfil
\subfloat[Joint delay and angle SAGE algorithm.]{\includegraphics[width=0.68\columnwidth]{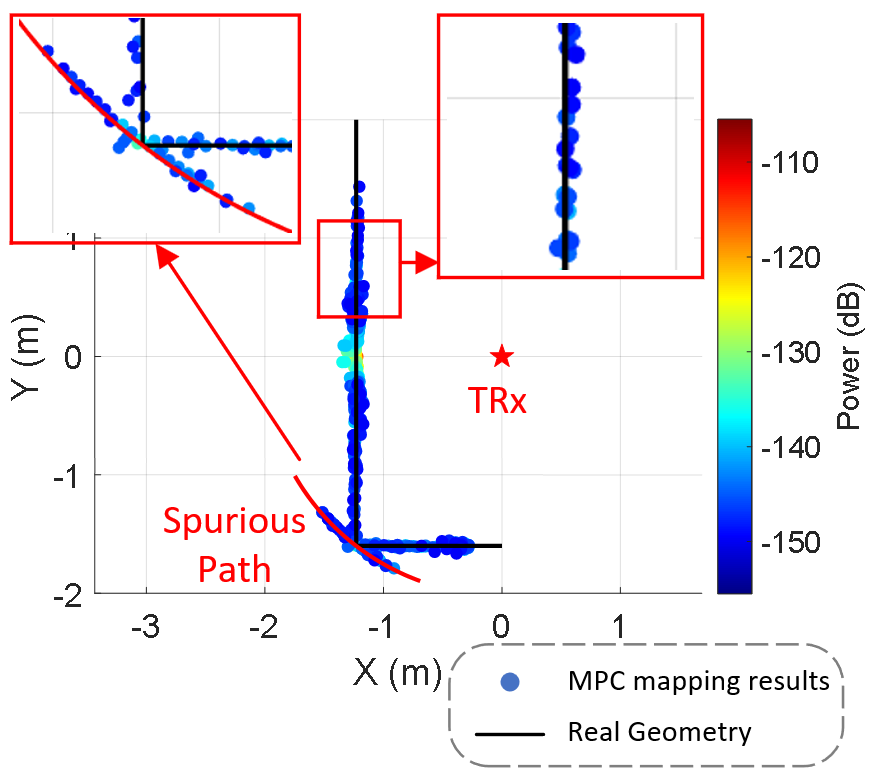}%
\label{fig:recon sage}}
\hfil
\subfloat[Joint delay and angle SAGE algorithm after removing spurious path.]{\includegraphics[width=0.68\columnwidth]{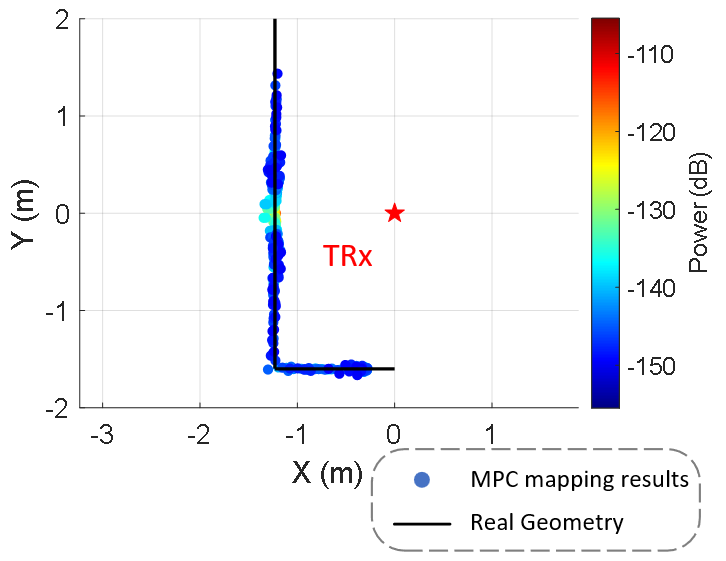}%
\label{fig:recon_MLE}}
\caption{Geometry reconstruction results based on three algorithms.}
\label{fig:corner}
\label{fig:recon_envir}
\end{figure*}

\begin{figure}[t]
\centering
\includegraphics[width=0.68\columnwidth]{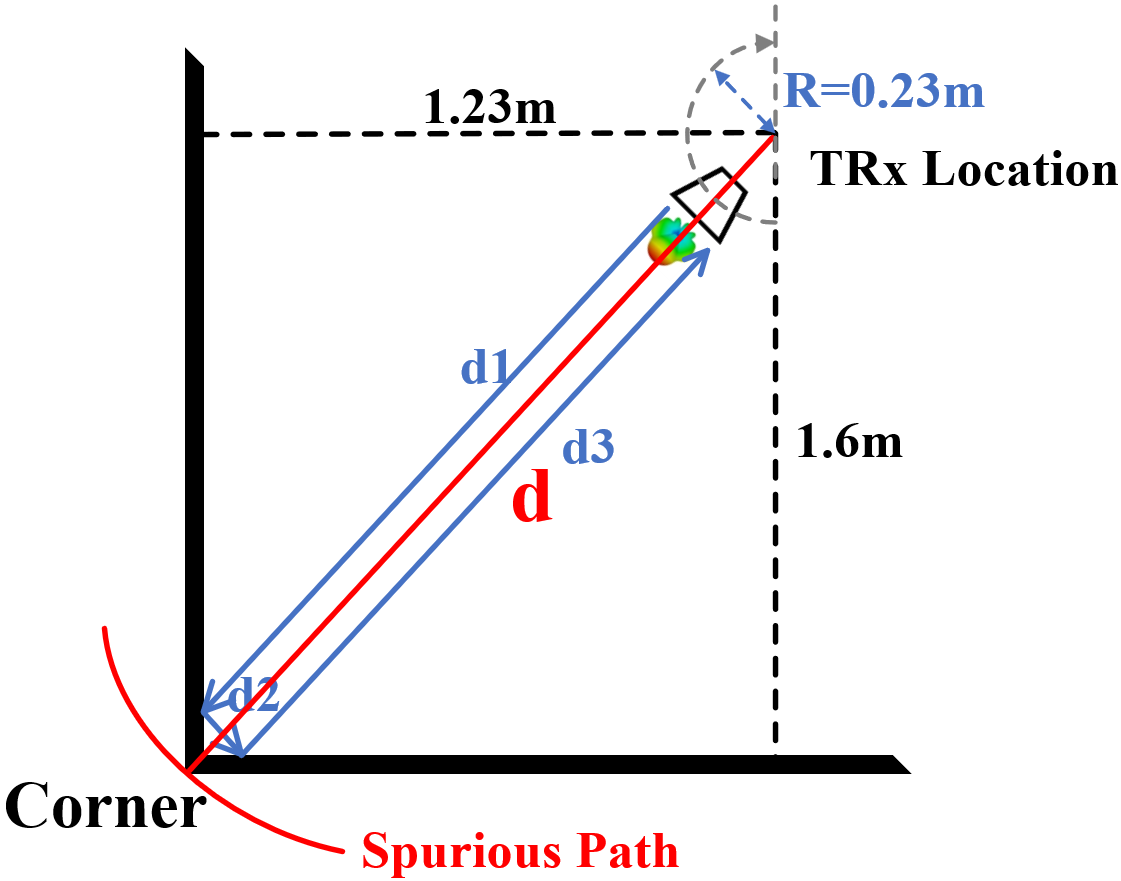}
\caption{The corner effect of spurious path.}
\label{fig:reflection}
\end{figure}
After estimating the parameters of each MPC, the corresponding mapping positions are calculated using geometric relationships. Specifically, the distance between TRx and the mapping point is determined by delay while the angle provides the direction information, which converts MPC parameters information into 2D-plane coordinates. For each MPC, the distance between the mapping point and the TRx in each sensing direction can be calculated as
\begin{equation}
    d_{E,\ell} = c\tau_{\ell}/2,
\end{equation}
where \( \tau_{\ell} \) is the estimated delay of the $\ell$-th MPC. Furthermore, the position of the mapping point can be calculated as
\begin{equation}
    \mathbf{r}_{\ell} = d_{E,\ell} \mathbf{\Omega}_\ell+\mathbf{r}_{S},
\end{equation}
where \( \mathbf{r}_{S} \) is the vector of the TRx location, which serves as the origin of the coordinate system. Moreover, \( \mathbf{\Omega}_\ell \) is the sensing direction vector in the azimuth domain, which is determined by the estimated angle of the $\ell$-th MPC.
\begin{equation}
    \mathbf{\Omega}_\ell = \left[ \cos \theta_\ell, \sin \theta_\ell \right]^T.
\end{equation}

Specifically, the spurious path from the corner is falsely detected by incorrect angle identification due to the existence of the antenna radiation pattern, as illustrated in Fig.~\ref{fig:reflection}. To address this issue, a geometry-based method is employed to eliminate the spurious path from corners. For second-order reflections occurring at the corner, the reflected and incident paths are nearly parallel, so the distance $d_2$ between two reflection points can be ignored as Tx and Rx are co-located at the same location. Assume $d$ is the direct reflection distance between the TRx and the corner. The total distance of the second-order reflection path is approximated as
\begin{equation}
    d_{\rm total} = d_1 + d_2 + d_3 \approx d_1 + d_3 \approx 2(d-R).
\end{equation}

The mapping points due to the spurious path can be modeled as a circular arc, with the TRx location as the center and the approximate distance $d_{\rm total}/2$ as the radius. By fitting this theoretical spurious path, points near this path can be removed from the mapping results, thus enhancing the accuracy of geometry reconstruction of corners.

\section{Geometry Reconstruction and Material Identification}
\begin{table}[t]
\caption{Ranging error of different methods.}
\label{tab:error}
\centering
\begin{tabular}{{ccc}}
\toprule
Method                               & MDE (cm)  & RMSE (cm) \\
\midrule
Maximum-search algorithm             & 2.76 & 4.93 \\
Joint delay and angle SAGE algorithm & 3.09 & 5.52 \\
Joint delay and angle SAGE algorithm                  & 1.75 & 2.53 \\
after removing spurious path & & \\
\bottomrule
\end{tabular}
\end{table}

\subsection{Geometry Reconstruction}

By calculating the geometry based on the extracted MPC parameters, the environment can be reconstructed, where the results are shown in Fig.~\ref{fig:recon_envir}. To compare, the commonly used maximum-search algorithm is implemented, which generates mapping points based on the echo signal parameters that are estimated as the parameters of the strongest path~\cite{li2023300}. As shown in Fig.~\ref{fig:recon_envir}(a), the algorithm captures the general geometry of the environment, however, it fails to detect the corner due to the existence of the antenna pattern. As shown in Fig.~\ref{fig:recon_envir}(b), the joint delay and angle SAGE algorithm in this work outperforms the maximum-search algorithm by considering the impact of the antenna radiation pattern and jointly estimating both delay and angle parameters, which results in a more precise geometry reconstruction. To further address the corner reconstruction issue, a geometry-based method introduced in section~\ref{sec:mapping} is employed to model the spurious path of the corner, as presented in Fig.~\ref{fig:recon_envir}(b). By removing the corner effect, more accurate reconstruction results are depicted in Fig.~\ref{fig:recon_envir}(c), which closely align with the real environment geometry.

\begin{figure}
\centering
\includegraphics[width=0.95\columnwidth]{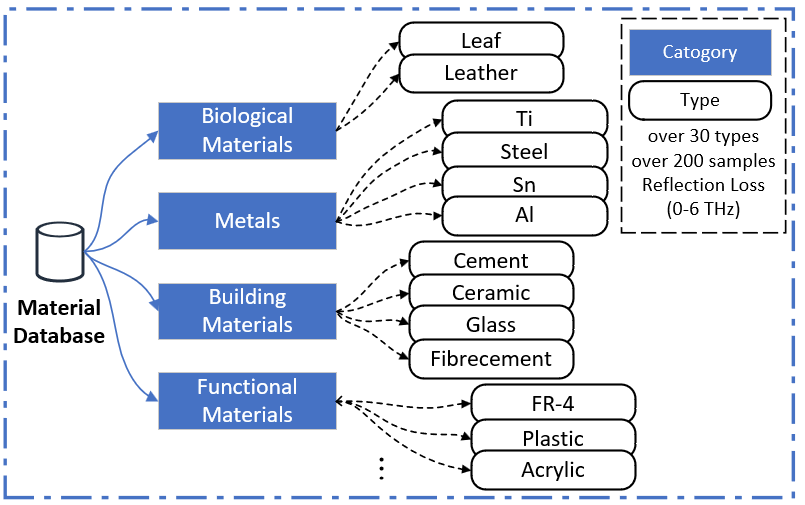}
\caption{Material Database measured by THz-TDS.}
\label{fig:database}
\end{figure}

\begin{table}[t]
\centering
\caption{Exemplary reflection loss of different materials at 300GHz}
\label{tab:loss}
\begin{tabular}{cc}
\toprule
Material      & Reflection loss at 300GHz (dB) \\
\midrule
Ti            & 0.84                          \\
Sn            & 1.72                          \\
Steel         & 2.42                          \\
Cement & 11.84                         \\
Ceramic       & 12.10                         \\
Fiber cement  & 13.09                         \\
Cardboard    & 16.86                         \\
Wood          & 20.42                         \\
\bottomrule
\end{tabular}
\end{table}
The ranging error of different algorithms is shown in Table~\ref{tab:error}. All methods achieve centimeter-level accuracy thanks to the ultrawide bandwidth, i.e., 20~GHz. The joint delay and angle SAGE algorithm after removing the spurious path achieves the lowest mean distance error (MDE) of 1.75~cm and root mean square error (RMSE) of 2.53~cm, demonstrating its superior capability in environment geometry reconstruction.

\subsection{Material Identification Based on THz-TDS}
The material reflection database is presented based on the measurements through THz-TDS, which initially captures the material response in the delay domain, and then is converted into the frequency domain using a discrete Fourier transform (DFT) function across 0-6~THz frequency range. A database comprising over 30 types and over 200 common material samples is built, as shown in Fig.~\ref{fig:database}. Reflection losses of several typical indoor materials at 300~GHz are presented in Table~\ref{tab:loss}, including steel, cement, wood, and so on.

Based on the channel measurement data, the maximum received power from the wall and window is selected for material identification, respectively. To calculate the reflection loss, the FSPL is subtracted from the maximum received power. For the wall, the reflection loss under vertical incidence conditions is calculated as 10.38~dB, which matches closely to the reflection loss of cement in the material database. Similarly, for the window, the reflection loss is calculated as 3.81~dB, which corresponds closely to the value of steel. The identification results are consistent with the materials of the wall and the metal structure of the window frame, respectively, validating the system's sensing ability to perform robust material identification based on a material database of reflection characteristics at 300~GHz.

\section{Conclusion}
In this paper, environment reconstruction experiments are conducted in an indoor scenario using a VNA-based sounder in 290–310~GHz band. The THz TRx scans the spatial domain using a mechanical rotator to perceive the surrounding environments. From the measured CIRs, a joint delay and angle SAGE algorithm is employed to estimate the delay and angle of MPCs jointly. Furthermore, by modeling and removing the spurious path from the corner caused by the antenna radiation pattern, the accuracy can achieve 1.75~cm. The 2D SAGE algorithm after removing the spurious path achieves high accuracy at the centimeter level and solves corner detection issue. Additionally, the reflection losses of the wall and window frame are calculated based on the measured echo signals. The results closely match cement and steel in our material database created through THz-TDS measurements, validating the accuracy of the material identification method.

\bibliographystyle{IEEEtran}
\bibliography{isac}
\end{document}